\documentclass[12pt]{article}                                                                                              
\usepackage{amssymb,amsmath}                                                                                               
\usepackage[noblocks]{authblk}                                                                                             
\usepackage{cite}                                                                                                          
\usepackage[top=0.75in, bottom=0.75in, left=0.75in, right=0.75in, dvips]{geometry}                                         
\usepackage{caption}                                                                                                       
\usepackage{multicol}                                                                                                      
\usepackage[pdftex]{graphicx}
\usepackage[export]{adjustbox}
\usepackage{epstopdf}
\usepackage{booktabs}
\usepackage{mwe}
\usepackage{leftidx}
\newcommand{\imineq}[2]{\vcenter{\hbox{\includegraphics[height=#2ex]{#1}}}}

\begin{document}                                                                                                           
\title{Perturbative Calculations with the First Order Form of Gauge Theories}
\author[1]{F. T. Brandt\footnote{fbrandt@usp.br}}                                                                          
\author[2]{D. G. C. McKeon\footnote{dgmckeo2@uwo.ca}}                                                                    

\affil[1]{Instituto de Fisica, Universidade de S$\mathrm{\tilde{a}}$o Paulo, S$\mathrm{\tilde{a}}$o Paulo, 
SP 05508-900, Brazil}                                                                                                                     
\affil[2]{Department of Applied Mathematics, University of Western Ontario, London Canada N6A 5B7}                         
\date{}                                                                                                                    
                                                                                                                           
\maketitle        

                                                                                                                           
\begin{abstract}
The first and second order form of gauge theories are classically equivalent; we consider the consequence of quantizing the first order
form using the Faddeev-Popov approach. Both the Yang-Mills and the Einstein-Hilbert actions are considered. An advantage of this approach
is that the interaction vertices are quite simple, being independent of momenta. It is necessary however to consider the propagator
for two fields (including a mixed propagator). We derive the Feynman rules for both models and consider the one loop correction for the thermal energy momentum tensor.
\end{abstract}

\section{Introduction}
Covariant quantization of classical Yang-Mills field only became possible when it was realized that non-physical modes of the vector field
had to be cancelled by contributions from so-called ``ghost''  fields that had non-trivial 
interactions \cite{Khriplovich:1969aa,Feynman:1963ax,DeWitt:1967uc,Faddeev:1967fc,Mandelstam:1968hz}.
Even then, computations are quite involved in large part because vertices arising from the classical second order Yang-Mills (2YM) Lagrangian
\begin{equation}\label{e1}
{\cal L}_{YM}^{(2)} = - \frac 1 4 \left(\partial_\mu A^a_\nu - \partial_\nu A^a_\mu + g f^{abc} A^b_\mu A^c_\nu\right)^2  
\end{equation}
are quite complicated; there is momentum dependent three point vertex as well as a four point vertex.

The second order Lagrangian of Eq. \eqref{e1} is classically equivalent to the first order Yang-Mills (1YM) Lagrangian
\begin{equation}\label{e2}
{\cal L}_{YM}^{(1)} = - \frac 1 2 F^a_{\mu\nu}\left(\partial^\mu A^{a\, \nu} - \partial^\nu A^{a\, \mu} + g f^{abc} A^{b\, \mu} A^{c\, \nu}\right)
+ \frac 1 4 F^a_{\mu\nu} F^{a\, \mu\nu}
\end{equation}
as once the equation of motion for the independent field $F^{a}_{\mu\nu}$ is used to eliminate it from the Lagrangian of Eq. \eqref{e2}, the Lagrangian of Eq. \eqref{e1} is recovered. The advantage of working directly with the Lagrangian of Eq. \eqref{e2} is that there is now only a relatively simple three point vertex $\operatorname{F-A-A}$. It is necessary however to work with not only propagators $\operatorname{A-A}$ and $\operatorname{F-F}$ for the fields
$A^{a}_{\mu}$ and $F^{a}_{\mu\nu}$ , but also a mixed propagator $\operatorname{A-F}$. 
This has been considered in \cite{McKeon:1994ds} using background field quantization.

The second order Einstein-Hilbert Lagrangian (2EH) written in terms of the metric is
\begin{equation}\label{e3}
{\cal L}_{EH}^{(2)} = -\kappa \sqrt{-g} g^{\mu\nu} R_{\mu\nu} (\Gamma) 
\end{equation}
where
\begin{equation}\label{e4}
R_{\mu\nu} = \Gamma^{\rho}_{\mu\rho,\nu}  - \Gamma^{\rho}_{\mu\nu,\rho} -  
\Gamma^{\sigma}_{\mu\nu}   \Gamma^{\rho}_{\sigma\rho} +  \Gamma^{\rho}_{\mu\sigma}   \Gamma^{\sigma}_{\nu\rho}   
\end{equation}
with
\begin{equation}\label{e5}
\Gamma^{\rho}_{\mu\nu} = \frac{1}{2} g^{\rho\lambda}\left(g_{\mu\lambda,\nu}+g_{\nu\lambda,\mu} - g_{\mu\nu,\lambda}\right)   .
\end{equation}
If we now set
\begin{eqnarray}\label{e67}
\sqrt{-g} g^{\mu\nu}  &=& h^{\mu\nu} \\
G^{\lambda}_{\mu\nu}   &=& \Gamma^{\lambda}_{\mu\nu} - \frac 1 2 \left(\delta^\lambda_\mu \Gamma^\sigma_{\nu\sigma} + \delta^\lambda_\nu \Gamma^\sigma_{\mu\sigma} \right)
\end{eqnarray}
then Eq. \eqref{e3} becomes
\begin{equation}\label{e8}
{\cal L}_{EH}^{(2)} = \kappa h^{\mu\nu} \left(G^\lambda_{\mu\nu,\lambda} + \frac{1}{d-1} G^\lambda_{\mu\lambda} G^\sigma_{\nu\sigma}   
- G^\lambda_{\mu\sigma} G^\sigma_{\nu\lambda}   \right),
\end{equation}
where $d$ is the space-time dimension.

The `` Faddeev-Popov'' (FP) quantization procedure of Refs. \cite{Feynman:1963ax,DeWitt:1967uc,Faddeev:1967fc,Mandelstam:1968hz} has been applied to the action
of Eq. \eqref{e3} with either $g_{\mu\nu}$ \cite{tHooft:1974bx,Goroff:1985sz,tHooft:2002xp} or $\sqrt{-g} g^{\mu\nu}$ \cite{Capper:1973pv,Capper:1974vb} 
being treated a being the independent field. (The FP has  to be extended to accomodate the ``transverse-traceless''  (TT) gauge \cite{Brandt:2007td}). Background
field quantization is employed \cite{Dewitt:1967ub,Honerkamp:1972fd,Abbott:1980hw} with $g_{\mu\nu}$ being expanded about a classical background
field such as the flat metric $\eta_{\mu\nu}$. This leads to exceedingly complicated vertices as $g$ and $g^{\mu\nu}$ now both become infinite
series in the quantum field. The part of ${\cal L}^{(2)}_{EH}$ that is just bilinear in the quantum field is a free second order spin-two Lagrangian.

If in Eqs. (\ref{e3},\ref{e4}), $g_{\mu\nu}$ and $\Gamma^\rho_{\mu\nu}$ are taken to be independent fields, then for $d > 2$ the equation of motion for
$\Gamma^\rho_{\mu\nu}$ results in Eq. \eqref{e5} \cite{Hobson:2006se}. (This was noted by Einstein  \cite{Einstein1925}; it is often a result credited to
Palatini \cite{Ferraris82}.) We will consider the first order Einstein-Hilbert (1EH) Lagrangian 
${\cal L}^{(1)}_{EH}$ being identical to ${\cal L}^{(2)}_{EH}$ in Eq. \eqref{e8} with $h^{\mu\nu}$ and $G^\lambda_{\mu\nu}$ being taken as independent fields. 
We then have only one relatively simple momentum independent vertex $\operatorname{G-G-h}$, with the propagators $\operatorname{h-h}$,
$\operatorname{G-G}$  and $\operatorname{h-G}$.

In $d=2$ dimensions, ${\cal L}^{(1)}_{EH}$ and ${\cal L}^{(2)}_{EH}$ are inequivalent; an extra vector field arises when solving the equation
of motion for $\Gamma^\lambda_{\mu\nu}$  \cite{Lindstrom:1987sg,Gegenberg:1987dw}. The Lagrangian ${\cal L}^{(2)}_{EH}$ in $d=2$ dimensions is not a total
divergence although its equations of motion are trivial and the constraint structure reveals that the gauge invariance is simply
$\delta g_{\mu\nu} = \epsilon_{\mu\nu} (x)$ for and arbitrary tensor $\epsilon_{\mu\nu}(x)$ \cite{Kiriushcheva:2005kk}. This shows that no physical
degrees of freedom reside in ${\cal L}^{(2)}_{EH}$ when $d=2$. When $d=2$, a canonical analysis of ${\cal L}^{(1)}_{EH}$ also possesses no physical degrees
of freedom but possesses an unusual local gauge invariance that is distinct from the manifest diffeomorphism invariance 
\cite{Kiriushcheva:2005tj,Kiriushcheva:2006gp}. Furthermore, upon quantizing ${\cal L}^{(1)}_{EH}$ when $d=2$ using the FP procedure, it
can be shown that all perturbative radiative effects vanish \cite{McKeon:2005be}.

We will now consider the quantization of ${\cal L}^{(1)}_{YM}$ and  ${\cal L}^{(1)}_{EH}$ when $d > 2$.

\section{First order Yang-Mills action}

The Lagrangian of Eq. \eqref{e2} is invariant under infinitesimal local gauge transformation
\begin{subequations}\label{e9}
\begin{equation}\label{e9a}
\delta A^a_\mu = D_\mu^{ab} \theta^b \equiv \left(\partial_\mu \delta^{ab} + g f^{apb} A^p_\mu \right) \theta^b
\end{equation}
\begin{equation}\label{e9b}
\delta F^a_{\mu\nu} =  a f^{apb} F_{\mu\nu}^p \theta^b
\end{equation}
\end{subequations}
necessitating introduction of a gauge fixing Lagrangian ${\cal L}_{gf}$ and its associated ghost Lagrangian 
${\cal L}_{gh}$ \cite{Khriplovich:1969aa,Feynman:1963ax,DeWitt:1967uc,Faddeev:1967fc,Mandelstam:1968hz}. Working  
with the covariant gauge fixing Lagrangian 
\begin{equation}\label{e10}
{\cal L}_{gf} =  -\frac{1}{2\alpha}\left(\partial\cdot A^a\right)^2
\end{equation}
one has
\begin{equation}\label{e11}
{\cal L}_{gh} =  {\bar c}^{\, a} \partial \cdot D^{ab} \,{c}^{\, b}
\end{equation}
where ${\bar c}^{a}$ and ${c}^{a}$ are the usual Fermionic scalar ghost fields.

\begin{figure}
\begin{eqnarray}
\displaystyle{\substack{{}^{\displaystyle{ F_{\lambda\sigma}^a}} {\; \includegraphics[scale=0.7]{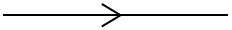} \;} {}^{\displaystyle{ F_{\rho\kappa}^b}} \\  {\displaystyle{ p}}}} &:& \;\;\;\;
2 i\left(I_{\lambda\sigma,\rho\kappa}-\frac{1}{p^2} L_{\lambda\sigma,\rho\kappa}(p)\right)\delta^{ab}
\nonumber 
\\
\nonumber 
\\
\displaystyle{\substack{ {}^{\displaystyle{A_{\mu}^a }}\; \includegraphics[scale=0.7]{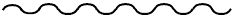} \; {}^{\displaystyle{ A_{\nu}^b }}\\  {\displaystyle{ p}}}} &:& \;\;\;\;
-\frac{i}{p^2}\left( \eta_{\mu\nu} -\frac{1-\alpha}{p^2}p_\mu p_\nu\right)\delta^{ab}
\nonumber 
\\
\nonumber 
\\
\displaystyle{\substack{{}^{\displaystyle{ A_{\mu}^a}} \; \includegraphics[scale=0.7]{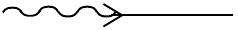} \; {}^{\displaystyle{ F_{\rho\kappa}^b}} \\  {\displaystyle{ p}}}} &:& \;\;\;\;
\frac{1}{p^2}\left(p_\rho \eta_{\kappa\mu} - p_\kappa\eta_{\rho\mu}\right)\delta^{ab}
\nonumber 
\\
\nonumber 
\\
\displaystyle{\substack{{}^{\displaystyle{F_{\lambda\sigma}^a}} \; \includegraphics[scale=0.7]{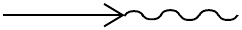} \; {}^{\displaystyle{ A_{\nu}^b}} \\  {\displaystyle{ p}}}} &:& \;\;\;\;
-\frac{1}{p^2}\left(p_\lambda \eta_{\sigma\nu} - p_\sigma\eta_{\lambda\nu}\right)\delta^{ab}
\nonumber
\\
\nonumber 
\\
F_{\lambda\sigma}^a { \imineq{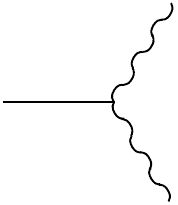}{14} }^{\displaystyle{\;\;A_\mu^b}}_{\displaystyle{\;\;A_\nu^c}}&:& \;\;\;\;
-\frac{i}{2}f^{abc}\left(\eta_{\lambda\mu} \eta_{\sigma\nu} - \eta_{\sigma\mu}\eta_{\lambda\nu}\right)
\nonumber
\\
\nonumber 
\\
\substack{ {}^{\displaystyle{c^a }}\; \includegraphics[scale=0.7]{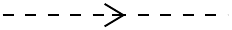} \; {}^{\displaystyle{ \bar c^b }}\\ 
 {\displaystyle{ p}}}&:& \;\;\;\;
\frac{i}{p^2}\delta^{ab}
\nonumber
\\
\nonumber 
\\
{ \displaystyle{A_\mu^a}  \atop {{\displaystyle{\bar c^b\;\; }}} {\includegraphics[scale=0.7]{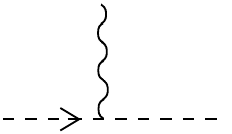}} {\displaystyle{c^c}}} 
\atop \displaystyle{p}\;\;\;\;\;\;\;\;\;\;\;&:& \;\;\;\;
-g f^{abc} p_\mu
\nonumber
\end{eqnarray}
\caption{Feynman rules for first order Yang-Mills}
\label{fig1}
\end{figure}

The terms in ${\cal L}^{(1)}_{YM}+{\cal L}_{gf}+{\cal L}_{gh}$ that are bilinear in the fields $A_\mu^a$ and $F_{\lambda\sigma}^a$ are
\begin{equation}\label{e12}
\frac{1}{2}\left(\begin{matrix}
A_\mu, & F_{\lambda\sigma}^a
\end{matrix}\right)
\left(\begin{matrix}
  \frac{1}{\alpha}\partial^\mu\partial^\nu & \frac{1}{2}\left(\partial^\rho\eta^{\kappa \mu} - \partial^\kappa \eta^{\rho\mu}\right) \\
 -\frac{1}{2}\left(\partial^\lambda\eta^{\sigma \nu} - \partial^\sigma \eta^{\lambda\nu}\right)  & 
\frac{1}{4}\left(\eta^{\lambda\rho}\eta^{\sigma \kappa} - \eta^{\lambda \kappa} \eta^{\sigma\rho}\right) 
\end{matrix}\right)
\left(\begin{matrix}
A_\nu \\ F_{\rho\kappa}^a
\end{matrix}\right).
\end{equation}
The inverse of the matrix appearing in Eq. \eqref{e12} is
\begin{equation}\label{e13}
\Delta(\partial) = 
\left(\begin{matrix}
\frac{1}{\partial^2}\left(\eta^{\mu\nu} -   \frac{(1-\alpha)}{\partial^2}\partial^\mu\partial^\nu\right) 
& -\frac{1}{\partial^2}\left(\partial^\rho\eta^{\kappa \mu} - \partial^\kappa \eta^{\rho\mu}\right) \\
 \frac{1}{\partial^2}\left(\partial^\lambda\eta^{\sigma \nu} - \partial^\sigma \eta^{\lambda\nu}\right)  & 
2\left(I^{\lambda\sigma, \rho \kappa} - \frac{1}{\partial^2}L^{\lambda \sigma,\rho\kappa}\right) 
\end{matrix}\right),
\end{equation}
where
\begin{subequations}\label{e14}
\begin{equation}\label{e14a}
I^{\lambda\sigma,\rho\kappa} = \frac{1}{2}\left(\eta^{\lambda\rho}\eta^{\sigma \kappa} - \eta^{\lambda \kappa} \eta^{\sigma\rho}\right) 
\end{equation}
\begin{equation}\label{e14b}
L^{\lambda\sigma,\rho\kappa}(\partial) = \frac{1}{2}\left(\partial^\lambda\partial^\rho \eta^{\sigma\kappa} 
+ \partial^\sigma\partial^\kappa \eta^{\lambda\rho} - \partial^\lambda\partial^\kappa \eta^{\sigma\rho} 
- \partial^\sigma\partial^\rho \eta^{\lambda\kappa} \right). 
\end{equation}
\end{subequations}
The propagators are given by $i\Delta(ip)$ and the $\operatorname{F-A-A}$ vertex follows from the interacting part of 
${\cal L}^{(1)}_{YM}$
\begin{equation}\label{e15}
-\frac{1}{2} g f^{abc} F^a_{\mu\nu} A^{b\, \mu} A^{c\, \nu}.
\end{equation} 
The Feynman rules appear in Fig. \ref{fig1}

We now turn to examining the 1EH Lagrangian.

\section{First order Einstein-Hilbert action}
It is tempting to consider directly applying the FP quantization procedure to the 1EH action of Eq. \eqref{e8} when $h^{\mu\nu}$ and
$G^\lambda_{\mu\nu}$ are treated as being independent fields. However, it is soon discovered that no choice of gauge leads to bilinears
in the effective Lagrangian that can be inverted so as to result in a suitable propagator. However, if we write
$h^{\mu\nu} = \eta^{\mu\nu} + \phi^{\mu\nu}$ where $\eta^{\mu\nu} = \mbox{diag}(+---\cdots -)$ is a flat background and
$\phi^{\mu\nu}$ is a quantum fluctuation, then Eq. \eqref{e8} becomes (with $\kappa = 1/2$)
\begin{eqnarray}\label{e16}
{\cal L}^{(1)}_{EH} &=& \frac{1}{2}\left[\phi^{\mu\nu} G^\lambda_{\mu\nu,\lambda} + 
\eta^{\mu\nu}\left(\frac{1}{d-1} G^\lambda_{\lambda\mu}G^\sigma_{\sigma\nu} - G^\lambda_{\sigma\mu}G^\sigma_{\lambda\nu}\right)\right]
\nonumber \\ &+&
\frac 1 2\left[\phi^{\mu\nu}\left(\frac{1}{d-1} G^\lambda_{\lambda\mu}G^\sigma_{\sigma\nu} - G^\lambda_{\sigma\mu}G^\sigma_{\lambda\nu}\right)\right]
\nonumber \\ &\equiv&
{\cal L}^{(1)\, 2}_{EH}  + {\cal L}^{(1)\, 3}_{EH} 
\end{eqnarray}

The infinitesimal form of diffeomorphism invariance associated with the action of Eq. \eqref{e8} is
\begin{subequations}\label{e17}
\begin{equation}\label{e17a}
\delta h^{\mu\nu} = h^{\mu\lambda} \partial_\lambda \theta^\nu + h^{\nu\lambda} \partial_\lambda \theta^\mu -\partial_\lambda(h^{\mu\nu}\theta^\lambda)
\end{equation}
\begin{eqnarray}\label{e17b}
\delta G^\lambda_{\mu\nu} &=& -\partial^2_{\mu\nu} \theta^\lambda
+\frac 1 2\left(\delta^\lambda_\mu\partial_\nu+\delta^\lambda_\nu\partial_\mu\right)\partial_\rho\theta^\rho-\theta^\rho\partial_\rho G^\lambda_{\mu\nu}
\nonumber \\ 
&+& G^\rho_{\mu\nu} \partial_\rho\theta^\lambda - \left(G^ \lambda_{\mu\rho} \partial_\nu+G^ \lambda_{\nu\rho} \partial_\mu\right)\theta^ \rho
\end{eqnarray}
\end{subequations}
which means that for ${\cal L}^{(1)}_{EH}$ in Eq. \eqref{e15} we have the gauge transformation of Eq. \eqref{e17b} while Eq. \eqref{e17a} now implies that
\begin{equation}\label{e18}
\delta\phi^{\mu\nu} = \partial^\mu\theta^\nu + \partial^\nu\theta^\mu 
+ \phi^{\mu\lambda}\partial_\lambda\theta^\nu + \phi^{\nu\lambda}\partial_\lambda\theta^\mu
- \eta^{\mu\nu}\partial\cdot\theta -\partial_\lambda(\phi^{\mu\nu}\theta^\lambda) .
\end{equation}
(Indices are now raised using $\eta^{\mu\nu}$.)

If we now choose the gauge fixing condition
\begin{equation}\label{e19}
{\cal L}_{gf} = -\frac{1}{2\alpha} (\partial_\mu\phi^{\mu\nu})^2
\end{equation}
then the Faddeev-Popov ghost contribution to the effective Lagrangian would be  \cite{Capper:1973pv,Capper:1974vb} 
\begin{eqnarray}\label{e20}
{\cal L}_{FP} &=& \bar d_\mu\left[ \partial^2 \eta^{\mu\nu} +(\partial_\rho\phi^{\rho\sigma})\partial_\sigma \eta^{\mu\nu}-
(\partial_\rho\phi^{\rho\mu})\partial^\nu \right. \nonumber \\
& &  \;\;\;\;\;\;\;\;\;\;\;  \left.  + \phi^{\rho\sigma} \partial_\rho\partial_\sigma \eta^{\mu\nu} -(\partial_\rho\partial^\nu\phi^{\rho\mu}) \right]d_\nu 
\end{eqnarray}

The terms bilinear in $\phi^{\mu\nu}$ and $G_{\mu\nu}^\lambda$ that follows from Eqs. \eqref{e16} and \eqref{e19} are
\begin{equation}\label{e21}
{\cal L}_{eff}^{(2)} =
\frac{1}{2}\begin{bmatrix}
\phi^{\mu\nu} , & G_{\alpha\beta}^\lambda 
\end{bmatrix}
\begin{bmatrix}
A_{\mu\nu\;\rho\kappa} & B_{\mu\nu\;\sigma}^{\gamma\delta} 
\\
C_{\rho\kappa\;\lambda}^{\alpha\beta}  & D^{\alpha\beta}_{\lambda} \; {}^{\gamma\delta}_{\sigma}  
\end{bmatrix}
\begin{bmatrix}
\phi^{\rho\kappa} \\ G_{\gamma\delta}^\sigma
\end{bmatrix},
\end{equation}
where
\begin{subequations}\label{eq22}
\begin{equation}\label{eq22a}
A_{\mu\nu\;\rho\kappa} \equiv \frac{1}{4\alpha}\left(
\partial_\mu\partial_\rho\eta_{\nu\kappa}+\partial_\nu\partial_\rho\eta_{\mu\kappa} +
\partial_\mu\partial_\kappa\eta_{\nu\rho}+\partial_\nu\partial_\kappa\eta_{\mu\rho}
\right)
\end{equation}
\begin{equation}\label{eq22b}
B_{\mu\nu\;\sigma}^{\gamma\delta}  \equiv\frac{1}{4}\left(
\delta^\gamma_\mu \delta^\delta_\nu + \delta^\gamma_\nu \delta^\delta_\mu
\right)\partial_\sigma
\end{equation}
\begin{equation}\label{eq22c}
C_{\rho\kappa\;\lambda}^{\alpha\beta}  \equiv -\frac{1}{4}\left(
\delta_\rho^\alpha \delta_\kappa^\beta + \delta_\rho^\beta \delta_\kappa^\alpha
\right)\partial_\lambda
\end{equation}
\begin{equation}\label{eq22d}
D^{\alpha\beta}_{\lambda} \; {}^{\gamma\delta}_{\sigma}   \equiv   
\frac{1}{4}\left[\left(
\frac{1}{d-1} \delta^{{\alpha}}_{{\lambda}} \delta^{{\gamma}}_{{\sigma}} \eta^{{\beta}{\delta}}
-\delta^{{\alpha}}_{{\sigma}} \delta^{{\gamma}}_{{\lambda}} \eta^{{\beta}{\delta}}
+ \alpha\leftrightarrow\beta\right) + \gamma\leftrightarrow\delta
\right]
\end{equation}
\end{subequations}

Using the blockwise matrix inversion
\begin{equation}\label{block}
\begin{bmatrix} \mathbf{A} & \mathbf{B} \\ \mathbf{C} & \mathbf{D} \end{bmatrix}^{-1} = \begin{bmatrix} 
\mathbf{X}^{-1} 
& - \mathbf{X}^{-1} \mathbf{BD}^{-1} \\ -\mathbf{D}^{-1}\mathbf{C} \mathbf{X}^{-1}  & 
\quad \mathbf{D}^{-1}+\mathbf{D}^{-1}\mathbf{C} \mathbf{X}^{-1} \mathbf{BD}^{-1}
\end{bmatrix}.
\end{equation}
where
\begin{equation}\label{X}  
\mathbf{X} = \mathbf{A}-\mathbf{BD}^{-1}\mathbf{C} 
\end{equation}
and $\mathbf{A}$, $\mathbf{B}$, $\mathbf{C}$ and  $\mathbf{D}$ have tensor representations given by Eqs. \eqref{eq22},
we can obtain the propagators in a straightforward way. (Some of the following steps were carried out using computer algebra.)
First, we compute the inverse of 
$D^{\alpha\beta}_{\lambda} \; {}^{\gamma\delta}_{\sigma}$. Using Eq. \eqref{eq22d}, we obtain
\begin{eqnarray}\label{Dm1}
D^{-1}{}^\lambda_{\alpha\beta} {}^\sigma_{\gamma\delta}  &=& \frac 1 2
\eta^{{\lambda}{\sigma}} \left(\eta_{{\alpha}{\gamma}}\eta_{{\beta}{\delta}}
+\eta_{{\alpha}{\delta}} \eta_{{\beta}{\gamma}}
-\frac{2} {d-2}\eta_{{\alpha}{\beta}}  \eta_{{\gamma}{\delta}} \right) 
 \nonumber \\ 
&-&\frac 1 2\left(  \delta^{{\lambda}}_{{\delta}}   \delta^{{\sigma}}_{{\beta}} \eta_{{\alpha}{\gamma}}
+ \delta^{{\lambda}}_{{\gamma}}  \delta^{{\sigma}}_{{\beta}} \eta_{{\alpha}{\delta}}
+  \delta^{{\lambda}}_{{\delta}} \delta^{{\sigma}}_{{\alpha}} \eta_{{\gamma}{\beta}}
+  \delta^{{\lambda}}_{{\gamma}} \delta^{{\sigma}}_{{\alpha}} \eta_{{\beta}{\delta}}\right).
\end{eqnarray}
Then, substituting Eqs. \eqref{eq22a}, \eqref{eq22b}, \eqref{eq22c} and \eqref{Dm1} into  the tensor form of Eq.  \eqref{X}, we obtain
($i\partial = p$)
\begin{eqnarray}\label{X1}
X_{\mu\nu\;\rho\kappa} &=&\frac{p^2}{8}\left(\frac{2 \eta_{\mu\nu}\eta_{\rho\kappa}}{d-2} 
-\eta_{\mu\rho}\eta_{\nu\kappa}  - \eta_{\mu\kappa}\eta_{\nu\rho}  \right)
\nonumber \\ 
&+& \left(\frac 1 8 - \frac{1}{4\alpha}\right)\left(
p_\mu p_\rho \eta_{\nu\kappa} + p_\nu p_\rho \eta_{\mu\kappa} 
p_\mu p_\kappa \eta_{\nu\rho} + p_\nu p_\kappa \eta_{\mu\rho}. 
\right)
\end{eqnarray}
Computing the inverse of this expression we obtain 
\begin{eqnarray}\label{props1}
X^{-1}{}^{\mu\nu\;\rho\kappa} &=&
\frac{1}{p^2}\left[(4-\alpha)\eta^{\mu\nu}\eta^{\rho\kappa} 
-2(\eta^{\mu\rho}\eta^{\nu\kappa}  + \eta^{\mu\kappa}\eta^{\nu\rho})  \right]
\nonumber \\ 
&+& \frac{\alpha-2}{p^4}\left[ 2\left(p^\mu p^\nu \eta^{\rho\kappa} + p^\rho p^\kappa\eta^{\mu \nu} \right) -
p^\mu p^\rho \eta^{\nu\kappa} - p^\nu p^\rho \eta^{\mu\kappa} -
p^\mu p^\kappa \eta^{\nu\rho} - p^\nu p^\kappa \eta^{\mu\rho} 
\right]
\nonumber \\ 
&\equiv & {\cal D}^{\phi^2}_{\mu\nu\;\rho\kappa},
\end{eqnarray}
where we have identified the result with the graviton propagator
${\cal D}^{\phi^2}_{\mu\nu\;\rho\kappa}$ (notice that for $\alpha=2$, ${\cal D}^{\phi^2}_{\mu\nu\;\rho\kappa}$
has the same structure as the DeDonder gauge propagator in the second order formulation).

Substituting Eqs. \eqref{eq22b}, \eqref{eq22c}, \eqref{Dm1} and \eqref{props1} into the tensor form of the off-diagonal blocks of Eq. \eqref{block} we obtain 
\begin{eqnarray}\label{Gp1}
{\cal D}^{G \phi}{}^\lambda_{\alpha\beta}{}^{\rho\kappa} & = &
\frac{i}{2p^2}\left[p_\alpha\left((\alpha -4) \delta^\lambda_\beta\eta^{\rho\kappa} 
+ 2 \delta^\rho_\beta\eta^{\lambda\kappa} + 2 \delta^\kappa_\beta\eta^{\lambda\rho} 
\right) -2 p^\lambda \delta_\alpha^\kappa \delta_\beta^\rho + \alpha \leftrightarrow \beta 
\right]
\nonumber \\
&-&\frac{i (\alpha -2)}{p^4}\left[
p^\kappa  p^\rho  ( p_\beta \delta^\lambda_\alpha  + p_\alpha \delta^\lambda_\beta )  
- p_\alpha p_\beta (p^\rho \eta^{\kappa\lambda}  + p^\kappa \eta^{\rho\lambda})  
+  p_\alpha p_\beta p^\lambda  \eta^{\kappa\rho} 
\right]
\end{eqnarray}
\begin{equation}\label{Gp2}
{\cal D}^{\phi G}{}^{\mu\nu} {}^\sigma_{\gamma\delta} =  - {\cal D}^{G \phi}{}^\sigma_{\gamma\delta}{}^{\mu\nu} 
\end{equation}
The propagator for the $G_{\mu\nu}^\lambda$ field can similarly be obtained computing the second diagonal block of \eqref{block} with the
help of Eqs. \eqref{eq22b}, \eqref{eq22c}, \eqref{Dm1} and \eqref{props1}, which yields
\begin{eqnarray}\label{GG1}
{\cal D}^{G^2}{}^\lambda_{\alpha\beta} {}^\sigma_{\gamma\delta} & = &
\frac{\alpha -2}{2 p^4}\left[
p_\alpha p_\beta p^\lambda\left( p_\delta \delta_\gamma^\sigma + p_\gamma \delta_\delta^\sigma \right) +
p_\gamma p_\delta p^\sigma\left( p_\alpha \delta^\lambda_\beta + p_\beta \delta_\alpha^\lambda \right) -
2 p_\alpha p_\beta p_\gamma p_\delta \eta^{\lambda\sigma}
\right]
\nonumber \\
& + & \frac{1}{4 p^2}\left[2 p^\lambda p^\sigma \left(\frac{2 \eta_{\alpha \beta} \eta_{\gamma\delta} }{d-2}
- \eta_{\alpha\gamma} \eta_{\beta\delta} - \eta_{\beta\gamma} \eta_{\alpha\delta}  \right)
\right. 
\nonumber \\ & &  \;\;\;\;\;\; 
+2 p^\lambda \left(p_\gamma\left( \delta_\alpha^\sigma \eta_{\beta\delta} + \delta_\beta^\sigma \eta_{\alpha\delta} \right)+
                           p_\delta\left( \delta_\alpha^\sigma \eta_{\beta\gamma} + \delta_\beta^\sigma \eta_{\alpha\gamma} \right) \right)
\nonumber \\ & &   \;\;\;\;\;\; 
+2 p^\sigma \left(p_\alpha\left( \delta_\gamma^\lambda \eta_{\beta\delta} + \delta_\delta^\lambda \eta_{\gamma\beta} \right)+
                           p_\beta\left( \delta_\gamma^\lambda \eta_{\delta\alpha} + \delta_\delta^\lambda \eta_{\alpha\gamma} \right) \right)
\nonumber \\ & &   \;\;\;\;\;\; 
-2 p_\alpha\left(p_\gamma\left(\eta^{\lambda\sigma} \eta_{\beta\delta} + \delta^\lambda_\delta \delta_\beta^\sigma\right) +
                      p_\delta\left(\eta^{\lambda\sigma} \eta_{\beta\gamma} + \delta^\lambda_\gamma \delta_\beta^\sigma\right)\right)
\nonumber \\ & &   \;\;\;\;\;\; 
-2 p_\beta\left(p_\gamma\left(\eta^{\lambda\sigma} \eta_{\alpha\delta} + \delta^\lambda_\delta \delta_\alpha^\sigma\right) +
                      p_\delta\left(\eta^{\lambda\sigma} \eta_{\alpha\gamma} + \delta^\lambda_\gamma \delta_\alpha^\sigma\right)\right)
\nonumber \\ & &   \;\;\;\;\;\; 
+(4-\alpha)\left(p_\gamma\delta_\delta^\sigma + p_\delta\delta_\gamma^\sigma\right) 
\left(p_\alpha\delta^\lambda_\beta+p_\beta\delta^\lambda_\alpha\right)  
\left. \frac{}{}\right]
\nonumber \\
&-&\frac{\eta^{\lambda\sigma}}{2}\left(\frac{2 \eta_{\alpha \beta} \eta_{\gamma\delta} }{d-2}
- \eta_{\alpha\gamma} \eta_{\beta\delta} - \eta_{\beta\gamma} \eta_{\alpha\delta} \right)
\nonumber \\
&-&\frac{\delta_\alpha^\sigma}{2}\left( \eta_{\beta \gamma}  \delta_\delta^\lambda + \eta_{\beta \delta}  \delta_\gamma^\lambda\right)
-\frac{\delta_\beta^\sigma}{2}\left( \eta_{\alpha \gamma}  \delta_\delta^\lambda + \eta_{\alpha \delta}  \delta_\gamma^\lambda\right)
\end{eqnarray}

There is only one interaction vertex which can be read from Eq. \eqref{e16}. The symmetrized result can be written as
\begin{eqnarray}\label{Gprop}
{\cal V} {}_{\mu\nu} {}^\lambda_{\alpha\beta}{}^\sigma_{\gamma\delta} & = & \frac{1}{8} \left\{\left[\left(
\frac{\eta_{\mu\beta} \eta_{\nu\delta} \delta_\alpha^\lambda \delta_\gamma^\sigma}{d-1} - 
\eta_{\mu\beta} \eta_{\nu\delta} \delta_\alpha^\sigma \delta_\gamma^\lambda + \mu \leftrightarrow \nu\right) + \alpha\leftrightarrow\beta
\right]+\gamma\leftrightarrow\delta\right\}
\end{eqnarray} 

The ghost propagator and vertex, which can be read in Eq. \eqref{e20}, are given by
\begin{equation}\label{ghprop}
{\cal D}^{gh}_{\mu\nu} = - \frac{\eta_{\mu\nu}}{p^2}
\end{equation}
and
\begin{equation}\label{ghvert}
{\cal V}^{gh}{\;\,}^{\mu\nu}_{\alpha\beta}(p_1,p_2,p_3) = \frac{\delta(p_1+p_2+p_3)}{2}\left[
\eta_{\alpha\beta}\left(p_2^\mu p_3^\nu + p_2^\nu p_3^\mu\right)
-p_2{}_\beta\left(p_1^\mu \delta_\alpha^\nu + p_1^\nu \delta_\alpha^\mu\right)
\right]
\end{equation}

\begin{figure}[h]
\begin{eqnarray}
\displaystyle{\substack{{}^{\displaystyle{ \phi^{\mu\nu}}} {\; \includegraphics[scale=0.7]{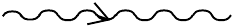} \;} {}^{\displaystyle{\phi^{\rho\kappa}}} \\  {\displaystyle{ p}}}} &:& \;\;\;\;
{\cal D}^{\phi^2}_{\mu\nu\;\rho\kappa},
\nonumber 
\\
\nonumber 
\\
\displaystyle{\substack{{}^{\displaystyle{G_{\alpha\beta}^\lambda}} \; \includegraphics[scale=0.7]{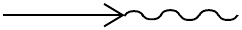} \; {}^{\displaystyle{ \phi^{\rho\kappa}}} \\  {\displaystyle{ p}}}} &:& \;\;\;\;
{\cal D}^{G \phi}{}^\lambda_{\alpha\beta}{}^{\rho\kappa} 
\nonumber
\\
\nonumber 
\\
\displaystyle{\substack{{}^{\displaystyle{ \phi^{\mu\nu}}} \; \includegraphics[scale=0.7]{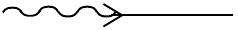} \; {}^{\displaystyle{ G_{\gamma\delta}^\sigma}} \\  {\displaystyle{ p}}}} &:& \;\;\;\;
{\cal D}^{\phi G}{}^{\mu\nu} {}^\sigma_{\gamma\delta} 
\nonumber 
\\
\nonumber 
\\
\displaystyle{\substack{{}^{\displaystyle{ G_{\alpha\beta}^\lambda}} {\; \includegraphics[scale=0.7]{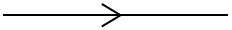} \;} {}^{\displaystyle{ G_{\gamma\delta}^\sigma}} \\  {\displaystyle{ p}}}} &:& \;\;\;\;
{\cal D}^{G^2}{}^\lambda_{\alpha\beta} {}^\sigma_{\gamma\delta} 
\nonumber 
\\
\nonumber 
\\
\phi_{\mu\nu} { \imineq{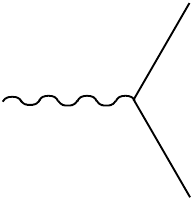}{14} }^{\displaystyle{\;\;G_{\alpha\beta}^\lambda}}_{\displaystyle{\;\;G_{\gamma\delta}^\sigma}}&:& \;\;\;\;
{\cal V} {}_{\mu\nu} {}^\lambda_{\alpha\beta}{}^\sigma_{\gamma\delta} 
\nonumber
\\
\nonumber 
\\
\substack{ {}^{\displaystyle{\bar d_\mu }}\; \includegraphics[scale=0.7]{ghost} \; {}^{\displaystyle{ d_\nu }}\\ 
 {\displaystyle{ p}}}&:& \;\;\;\;
{\cal D}^{gh}_{\mu\nu} 
\nonumber
\\
\nonumber 
\\
{ \displaystyle{\phi^{\mu\nu}}  \atop {{\displaystyle{\bar d_\beta\;\; }}} {\includegraphics[scale=0.7]{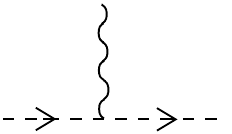}} {\displaystyle{d_\alpha}}} 
\atop \displaystyle{p_3}\;\;\;\;\;\;\;\;\;\;\; \displaystyle{p_2} &:& \;\;\;\;
{\cal V}^{gh}{\;\,}^{\mu\nu}_{\alpha\beta}(p_1,p_2,p_3) 
\nonumber
\end{eqnarray}
\caption{Feynman rules for first order Gravity.}
\label{fig2}
\end{figure}

Using Eqs. \eqref{props1}, \eqref{Gp1}, \eqref{Gp2}, \eqref{GG1}, \eqref{Gprop}, \eqref{ghprop} and \eqref{ghvert} we put together in Fig.
\ref{fig2} all the Feynman rules for first order gravity.

As an example of the effectiveness of the perturbative first order formalism,  
let us now consider an explicit perturbative calculation which makes use of the Feynman rules in Fig. \ref{fig2}.
We will consider a simple one-loop calculation which takes into account the coupling of 
the graviton field to the energy momentum tensor of a thermal gravitational plasma.
Since this is a well known result which has been obtained in the usual formulation of 
thermal gravity \cite{Gribosky:1988yk} as well as in the transverse traceless gauge fixing formulation 
\cite{Brandt:2007td}, it provides a simple test of the consistence of the first order formalism.

\begin{figure}
\[
  {\mbox{\includegraphics[scale=0.7]{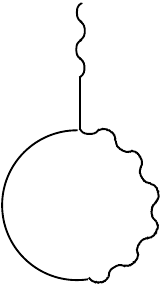}}  \atop \mbox{(a)} }\qquad
  {\mbox{\includegraphics[scale=0.7]{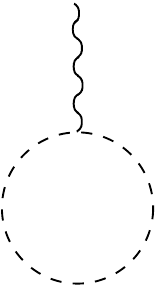}}  \atop \mbox{(b)} }\qquad
  {\mbox{\includegraphics[scale=0.7]{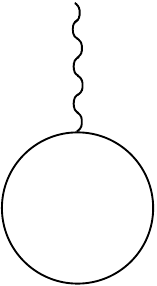}}  \atop \mbox{(c)} }\qquad
\]
\caption{Diagrams which contributes to the thermal energy momentum tensor.}
\label{fig3}
\end{figure}

The energy momentum tensor $T_{\mu\nu}$ and the one-graviton function $\Gamma_{\mu\nu}$ are related by  
\begin{equation}
\Gamma_{\mu\nu}=\frac{\delta \Gamma}{\delta \phi^{\mu\nu}} = -\frac{1}{2}\sqrt{-g}\,  T_{\mu\nu}  ,
\end{equation}
where $\Gamma$ is the one-loop thermal effective action. 
In the figure \ref{fig3} we shown the one-loop diagrams which contribute to $\Gamma^{\mu\nu}$. 
Using the imaginary time formalism \cite{kapusta:book89} the thermal part of each of these diagrams can be written as
\begin{eqnarray}\label{TermVac}
\int\,\frac{{\rm d}^{d-1}  k}{(2\pi)^{d-1}}
\int_{-i\infty+\delta}^{i\infty+\delta}
\frac{{\rm d} k_0}{2\pi i} N_B(k_0)
\left[f^I _{\mu\nu} (k)+f^I_{\mu\nu}(- k)\right],
\end{eqnarray}
where $N_B(k_0)$ is the Bose-Einstein thermal distribution function. For convenience we are considering the more general case of a 
$d$-dimensional space-time. The integrand of each contribution from Fig. \ref{fig3}  is denoted by 
$f^I_{\mu\nu}(k)$ ($I=a,b,c$).

Let us first consider the diagram with a mixed propagator as shown in Fig. \ref{fig3}-(a). 
Using the Feynman rules in Fig. \eqref{fig2}  we obtain
\begin{equation}
f^a_{\mu\nu}(k) = - i \left[(\alpha -2)  \frac{k^\lambda k_\alpha k_\beta}{k^4} 
-\frac{(3\alpha+2 d -4)k^\lambda \eta_{\alpha\beta} - d (k_\alpha \delta_\beta^\lambda +k_\beta \delta_\alpha^\lambda) }{2k^2} 
\right] (D B^{-1} X)^{\alpha\beta}_{\lambda\;\mu\nu},
\end{equation}
where the factor $ (D B^{-1} X)^{\alpha\beta}_{\lambda\;\mu\nu}$ produces the corresponding amputated Green function.
Since $f^a_{\mu\nu}(k)$ is an odd function of $k$ the net result in Eq. \eqref{TermVac} will vanish trivially. 

We are then left with the ghost loop and the $G$-loop contributions.  From Figs \ref{fig3}-(b) and  \ref{fig3}-(c) we obtain
\begin{equation}\label{fb1}
f^b_{\mu\nu}(k) = -d \frac{k_\mu k_\nu}{k^2}
\end{equation}
and
\begin{equation}\label{fc1}
f^c_{\mu\nu}(k) = \frac{1}{4}\left[ d(d-1) \eta_{\mu\nu} + d(d+1) \frac{k_\mu k_\nu}{k^2}
\right].
\end{equation}
Since we are using dimensional regularization, the first term in Eq. \eqref{fc1} produces a vanishing contribution when inserted into \eqref{TermVac}. Adding the non-vanishing contribution from $f^b_{\mu\nu}(k)$ and $f^c_{\mu\nu}(k)$, and using \eqref{TermVac} we obtain
\begin{equation}
\Gamma^{therm}_{\mu\nu} = \frac{d(d-3)}{2}\int\,\frac{{\rm d}^{d-1}  k}{(2\pi)^{d-1}}
\int_{-i\infty+\delta}^{i\infty+\delta} \frac{{\rm d} k_0}{2\pi i} N_B(k_0) \frac{k_\mu k_\nu}{k^2},
\end{equation}
where the factor $d(d-3)/2$ counts the degrees of freedom of a graviton in $d$ dimensions.
Closing the contour of integration in the right hand side plane, the pole at $k_0=|\vec k|$ gives the following contribution
[there is a minus sign from the clockwise contour integration and the pole from $1/k^2$ at $k_0=|\vec k|$ yields a factor
$1/(2 |\vec k|$)] 
\begin{eqnarray}
\Gamma^{therm}_{\mu\nu} &=& -\frac{d(d-3)}{4}
\int_0^\infty {\rm d} |\vec k|  \frac{|\vec k|^{d-1} }{{\rm e}^{\frac{|\vec k|}{T}} -1} 
\int\,\frac{{\rm d} \Omega_{d-1}  }{(2\pi)^{d-1}} \hat k_\mu \hat k_\nu
\nonumber \\ & = &
-\frac{d(d-3)}{4}  \zeta(d) \Gamma(d) T^d
\int\,\frac{{\rm d} \Omega_{d-1}  }{(2\pi)^{d-1}} \hat k_\mu \hat k_\nu,
\end{eqnarray}
where $\hat k_\mu = (1,\vec k/|\vec k|)$.  This result can be expressed in terms of the heat bath four-velocity $u_\mu=(1,0)$ as follows
\begin{eqnarray}
\Gamma^{therm}_{\mu\nu} &=& 
\frac{d(d-3)}{4(d-1)}  \zeta(d) \Gamma(d) \left(\int\,\frac{{\rm d} \Omega_{d-1}  }{(2\pi)^{d-1}} \right) T^d (\eta_{\mu\nu}- d u_\mu u_\nu)
\nonumber \\ & = &
\frac{d(d-3)}{4(d-1)}  \zeta(d) \Gamma(d)  \frac{2 \pi^{\frac{d-1}{2}}}{\Gamma\left(\frac{d-1}{2}\right)} \frac{T^d}{(2\pi)^{d-1}}  
(\eta_{\mu\nu}- d u_\mu u_\nu).
\end{eqnarray}
For $d=4$ we obtain
\begin{equation}
\left.\Gamma^{therm}_{\mu\nu}\right|_{d=4} =  \frac{\pi^2 T^4}{90} \left(\eta_{\mu\nu}-  4 u_\mu u_\nu\right),
\end{equation}
which is in agreement with the known result obtained using the second order formalism \cite{Gribosky:1988yk}.

\section{Discussion}
We have examined how the first order form of both the Yang-Mills and Einstein-Hilbert action can be used to compute quantum effects. In both cases, using the first order form at the action simplifies the vertices encountered when using the Faddeev-Popov quantization; unfortunately the propagators become more involved.

The first and second order form of the actions can be shown to be classically equivalent by examining the classical equations of motion. To show that the path integrals associated with ${\cal L}^{(2)}_{YM}$ and ${\cal L}^{(1)}_{YM}$ are equivalent, we need only take 
\begin{equation}\label{e24}
{\cal L}^{(2)}_{eff} = {\cal L}^{(2)}_{YM}  + {\cal L}_{gf}  + {\cal L}_{gh} 
\end{equation}
using Eqs. \eqref{e1}, \eqref{e10} and \eqref{e11} and insert into the path integral
\begin{equation}\label{e25}
{Z}^{(2)}_{eff} = \int {\cal D} A^a_\mu {\cal D} c^a {\cal D} \bar c^a  \exp i\int dx {\cal L}^{(2)}_{eff}
\end{equation}
the constant
\begin{equation}\label{e26}
\int {\cal D} F^a_{\mu\nu} \, \exp \,i\int dx\left(\frac 1 4 F^a_{\mu\nu} F^{a\,\mu\nu}\right).  
\end{equation}
Upon performing the shift
\begin{equation}\label{e27}
 F^a_{\mu\nu} \rightarrow F^a_{\mu\nu} - \left(\partial_\mu A_\nu -\partial_\nu A_\mu + g f^{abc} A^b_\mu A^c_\nu\right)  
\end{equation}
we convert ${Z}^{(2)}_{eff}$ into ${Z}^{(1)}_{eff}$ where
\begin{equation}\label{e28}
{Z}^{(1)}_{eff} = \int {\cal D} A^a_\mu  {\cal D} F^a_{\mu\nu} {\cal D} c^a {\cal D} \bar c^a  \exp i\int dx {\cal L}^{(1)}_{eff}
\end{equation}
where ${\cal L}^{(1)}_{eff}$ is identical to ${\cal L}^{(2)}_{eff}$ of Eq. \eqref{e24} except that now ${\cal L}^{(1)}_{YM}$ of Eq. \eqref{e2} replaces ${\cal L}^{(2)}_{YM}$.

Unfortunately, it is not so straightforward to show that when the Faddeev-Popov quantization procedure is used in conjunction with ${\cal L}^{(1)}_{EH}$, the same result is obtained as when ${\cal L}^{(2)}_{EH}$ is treated this way. In any case, it is not clear that the Faddeev-Popov procedure is appropriate for ${\cal L}^{(1)}_{EH}$ as the constraint structure of this Lagrangian implies that the functional measure receives a non-trivial contribution from second class constraints \cite{Chishtie:2013fna}.
Such contributions have also a significant effect when quantizing a model with an antisymetric tensor field interacting with a non-Abelian vector field and possesses a pseudoscalar mass \cite{Chishtie:2013rva}. 

The problem of renormalizing the divergences that arise when using the Faddeev-Popov approach to quantizing ${\cal L}^{(1)}_{YM}$ and ${\cal L}^{(1)}_{EH}$ is quite delicate on account of the presence of mixed propagators. We are currently considering this issue.

\noindent
{\bf Acknowledgments}

\noindent
We would like to thank CNPq and Fapesp (Brazil) for financial support. Roger Macleod helped with this work.


\end{document}